\documentstyle[aps,epsf,epsfig]{revtex}
\def \be   {\begin{equation}}
\def \ee   {\end{equation}}
\def \l {\label}
\begin{document}
\input epsf
\baselineskip=25pt
\title{DISCRETE CLASSICAL ELECTROMAGNETIC FIELDS}
\author{Manoelito M de Souza}
\address{Universidade Federal do Esp\'{\i}rito Santo - Departamento de
F\'{\i}sica\\29065.900 -Vit\'oria-ES-Brasil}
\date{\today}
\maketitle
\begin{abstract}
\noindent 
The classical electromagnetic field of a spinless point electron is described in a formalism with extended causality by discrete finite transverse point-vector fields with discrete and localized point interactions.  These fields are taken as a classical representation of photons, ``classical photons". They are all transversal photons; there are no scalar nor longitudinal photons as these are definitely eliminated by the gauge condition. The angular distribution of emitted photons coincides with the directions of maximum emission in the standard formalism. The Maxwell formalism and its standard field are retrieved by the replacement of these discrete fields by their space-time averages, and in this process scalar and longitudinal photons are necessarily created and added. Divergences and singularities are by-products of this averaging process. This formalism enlighten the meaning and the origin of the non-physical photons, the ones that violate the Lorentz condition in manifestly covariant quantization methods.
\end{abstract}
\begin{center}
\end{center}

The concept of extended causality \cite{hep-th/9610028,hep-th/9610145,hep-th/9708066,hep-th/9708096} allows a formalism for field theory in terms of discrete point-like fields with localized point-interactions and free of singularities and divergences. This is relevant for pointing the way to a finite description of fundamental fields in terms of point-like objects, a possible alternative to string motivated approaches. In this note we write Classical Electrodynamics in this formalism for the description of the field of a classical spinless point electron. We start, for completeness, with a brief review of reference \cite {hep-th/9708096}, about extended causality and its applications to field theory, before showing how it applies to the electromagnetic field.\\
In the standard field theory formalism the evolution of a field is constrained by
\be
\label{1}
\Delta\tau^2=-\Delta x^{2},
\ee 
where $\Delta\tau$ is the change of proper time associated to $\Delta x$. Geometrically (\ref{1}) is the definition of a three-dimensional double cone in a Minkowski spacetime of metric $\eta=diag(1,1,1,-1);$ $\Delta x$ is the four-vector separation between a generic event $x^{\mu}\equiv({\vec x},t)$ and the cone vertex. See the Figure 1.  This hypersurface is the support manifold for the field definition: a free field cannot be inside nor outside but only on the cone. The cone-aperture angle $\theta$ is given by
\be
\l{theta}
\tan\theta=\frac{|\Delta {\vec x}|}{|\Delta t|},\qquad c=1.
\ee
A change of cone-support corresponds to a change of speed and would be an indication of interactions.
Special Relativity restricts $\theta$ to the range $0\le\theta\le\frac{\pi}{4},$ which corresponds to a restriction on $\Delta\tau:$
\be
0<|\Delta\tau|\le|\Delta t|.
\ee
\parbox[]{7.5cm}{
\begin{figure}
\hspace{-5.0cm}
\epsfxsize=400pt
\epsfbox{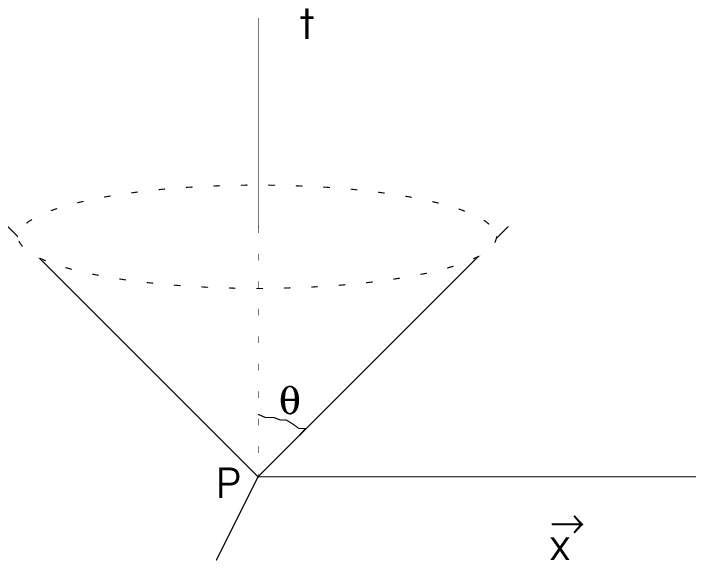}
\vglue-5cm
\end{figure}
\vglue-8cm%
}
{}\\
\mbox{}
\hfill
\hspace{5.0cm}
\parbox[]{7.5cm}{\vglue-5cm Fig. 1.
The relation $\Delta\tau^2=-\Delta x^{2},$ a causality constraint, is seen as a restriction of access to regions of spacetime. It defines a three-dimension cone which is the spacetime available to a point physical object at the cone vertex.}\\ \mbox{}
\vglue-1cm

This is local causality. The concept of extended causality corresponds to a more restrictive constraint; it requires that (\ref{1}) be applied at two neighboring events x and $x+dx$. This is equivalent to the imposition of a second constraint, besides the first one (\ref{1}):
\be
\l{f}
\Delta\tau+  f.\Delta x=0,
\ee
where $f$ is defined by $f^{\mu}=\frac{dx^{\mu}}{d\tau},$ a constant  four-vector tangent to the cone; it is  spacelike $(f^{2}=-1$) if $\Delta\tau\ne0,$  or timelike $(f^{2}=0$) if $\Delta\tau\ne0$. \\
The equation (\ref{f}) can be obtained from direct differentiation of (\ref{1}), $\Delta x.dx+\Delta\tau d\tau=0$, and geometrically (\ref{f}) defines a hyperplane tangent to the cone (\ref{1}). We have from (\ref{f}) that $$f_{\mu}=-\frac{\partial\tau}{\partial x^{\mu}},$$ where $\tau$ is a solution of (\ref{1}) and is seen as a function of x: $\tau=\tau_{0}\pm\sqrt{-(\Delta x)^{2}}$.  For $\Delta\tau=0,\;$ $f_{\mu}$ is orthogonal to the hyperplane (\ref{f}).\\ Imposing in field theory the two constraints, (\ref{1}) and (\ref{f}), instead of just (\ref{1}), as it is usually done, correspond to knowing the initial position and velocity in point-particle dynamics.

Together, the constraints (\ref{1}) and (\ref{f}) are equivalent to the single condition
\be
\l{lambda}
\Delta x.\Lambda^{f}.\Delta x=0,
\ee
with $\Lambda^{f}_{\mu\nu}=\eta_{\mu\nu}+f_{\mu}f_{\nu},$ ($f_{\mu}=\eta_{\mu\nu}f^{\nu}$),  which is a projector orthogonal to  $f^{\mu}$,  $  f.\Lambda.f=0.$ Therefore the constraint (\ref{lambda})  only allows displacements $\Delta x^{\mu}$ that are parallel to $f^{\mu}.$  A fixed four-vector $f$ at a point represents a fiber in the spacetime, a straight line tangent to $f^{\mu}$, the $f$-generator of the local cone (\ref{1}).
For a massless field (\ref{1}) defines a lightcone, (\ref{f}) defines a hyperplane tangent to this lightcone, and $f^{\mu}$ is the lightcone generator tangent to this hyperplane.\\
$A_{f}(x)$ is a f-field, that is, a field defined on a fiber   f. It is distinct of the field $A(x)$ of the standard formalism, which is defined on the entire cone. $A_{f}(x)$ may be seen as the restriction of $A(x)$ to a fiber $f$ 
\be
\l{Af}
A(x,\tau)_{f}=A(x,\tau){\Big |} _{dx.\Lambda^{f}.dx=0}
\ee
It is a point-like field, the intersection of the wave-front $A(x,\tau)$ with the fiber $f$. 
Conversely, we have the following relation 
\be
\label{s}
A(x,\tau)=\frac{1}{4\pi}\int d^{2}\Omega_{f}A_{f}(x,\tau),
\ee
where the integral represents the sum over all $f$ directions on a cone. $4\pi$ is a normalization factor. The physical interpretation associates $A_{f}(x)$, a point-perturbation propagating along the lightcone generator $f,$ with a physical photon - it is called a classical  photon - and $A(x)$, the standard field, to the effect of the exchanged classical photon averaged in time and in space. The Figure 2 shows the relationship between the fields $A_{f}$ and $A$ for a process involving the emission of a single physical photon $A_{f}(x)$; $A$ is its space average.

The derivatives of $A_{f}(x),$ allowed by the constraint (\ref{lambda}), are the directional derivatives along $f,$ which with the use of (\ref{f}) or of (\ref{lambda}) we write as
\be
\label{fd}
\partial_{\mu}A_{f}=(\frac{\partial }{\partial x^{\mu}}+\frac{\partial \tau}{\partial x^{\mu}}\frac{\partial}{\partial \tau})A(x,\tau){\Big |} _{dx.\Lambda^{f}.dx=0}={\Big(}\frac{\partial }{\partial x^{\mu}}-f_{\mu}\frac{\partial}{\partial \tau}{\Big)}A_{f}\equiv\nabla_{\mu} A_{f}.
\ee
With $\nabla$ replacing $\partial$ for taking care of the constraint (\ref{lambda}), the propertime $\tau$ is treated as a fifth independent parameter or coordinate. 

\parbox[]{7.5cm}{
\begin{figure}
\vglue-6cm
\epsfxsize=400pt
\epsfbox{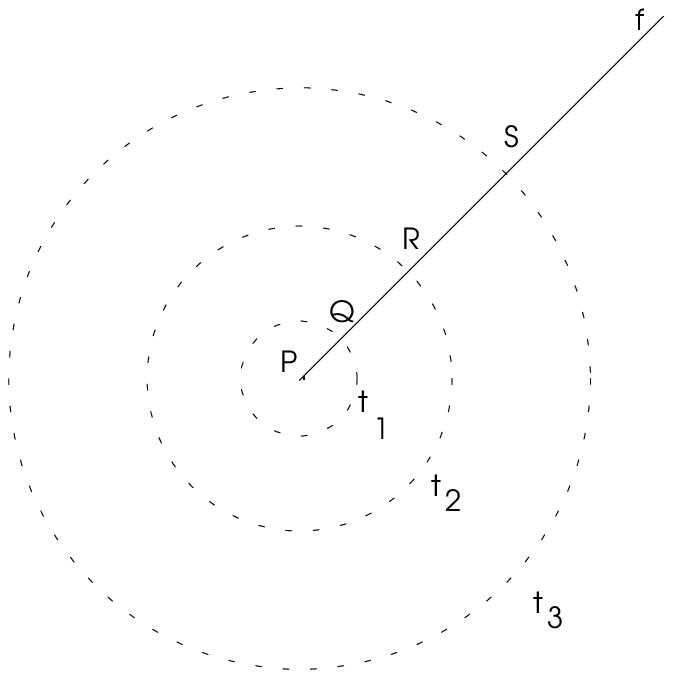}
\vglue-10cm
\end{figure}
\vglue-0cm}\\
\mbox{}
\hfill
\hspace{5.0cm}
\parbox[]{7.5cm}{\vglue-1cm Fig. 2.
The relationship between the fields $A_{f}$ and $A$. The three doted circles represent, at three instants of time, the field $A$ as an spherically symmetric signal emitted by a charge at the point P. The straight line PQRS\dots is the fiber $f,$ a lightcone generator tangent to $f^{\mu}.$ The points Q, R, and S are a classical photon $A_{f}$ at three instants of time.}\\ \mbox{}

The field equation for a massless field is
\be
\label{wef}
\eta^{\mu\nu}\nabla_{\mu}\nabla_{\nu}A_{f}(x,\tau)=J(x,\tau),
\ee
or, explicitly
\be
\label{wef'}
(\eta^{\mu\nu}\partial_{\mu}\partial_{\nu}-2f^{\mu}\partial_{\mu})A_{f}(x,\tau)=J(x,\tau),
\ee
as $f^{2}=0$. $J$ is its source four-vector current.\\ 
An integration over the $f$ degrees of freedom in (\ref{wef}) reproduces, with the use of (\ref{s}), the usual wave equation of the standard formalism, \be
\l{10'}
\eta^{\mu\nu}\partial_{\mu}\partial_{\nu} A(x)=4\pi J(x),
\ee
 as $\int d^{2}\Omega_{f}f^{\mu}\partial_{\mu}\partial_{\tau}A_{f(x)}=0$ because \cite{hep-th/9708066} $A_{f}(x)=A_{-f}(x)$. Observe that the missing $4\pi$ in (\ref{wef}) and in (\ref{wef'}) has re-appeared in (\ref{10'}) after using (\ref{s}) to retrieve the standard wave equation of A.\\ So, the standard formalism is retrieved from this $f$-formalism with the $A(x)$ as the average of $A_{f}(x)$, in the sense of (\ref{s}). $A_{f}(x)$ can be seen as a ``photon", a classical photon,  a particle-like classical field of which the  electromagnetic field $A(x)$ represents just an average, as the eq. (\ref{s}) can be interpreted.  We shall expose next the structure of Classical Electrodynamics written in terms of $A_{f}(x).$\\
\begin{center}
The Green's function
\end{center}

The $f$-wave equation (\ref{wef}) can be solved by a f-Green's function,
\be
\label{sgf}
A_{f}(x,\tau_{x})=\int d^{4}yd\tau_{y}\; G_{f}(x-y,\tau_{x}-\tau_{y})\;J(y),
\ee
with $G_{f}(x-y,\tau_{x}-\tau_{y})$ being a solution of
\be
\label{gfe}
\eta^{\mu\nu}\nabla_{\mu}\nabla_{\nu}G_{f}(x-y,\tau_{x}-\tau_{y})=\delta^{4}(x-y)\delta(\tau_{x}-\tau_{y}):=\delta^{5}(x-y).
\ee
This equation has been solved in \cite{hep-th/9708066}:
\be
\label{pr9}
G_{f}(x,\tau)=\frac{1}{2}\theta(-b{\bar f}.x)\theta(b\tau)\delta(\tau+  f.x);
\ee
where $b =\pm1,$ and $\theta (x)$ is the Heaviside function, $\theta(x>0)=1$ and $\theta(x<0)=0.$  $G_{f}(x,\tau)$ does not depend on ${\vec x}_{\hbox {\tiny T}}$, where the subscript ${\hbox {\tiny T}}$ stands for transversity with respect to ${\vec f},$: ${\vec f}.{\vec x}_{{\hbox {\tiny T}}}=0;$  this corresponds to an implicit $\delta^{2}({\vec x}_{\hbox {\tiny T}})$ in (\ref{pr9}).\\
The properties of $G_{f}$ are discussed in \cite{hep-th/9708066}. For $f^{\mu}=({\vec f}, f^{4})$, ${\bar f}$ is defined by ${\bar f}^{\mu}=(-{\vec f}, f^{4});$ $f$ and ${\bar f}$ are two opposing generators of a same lightcone; they are associated, respectively, to the $b=+1$ and to the $b=-1$ solutions and, therefore, to the processes of creation and annihilation of a photon.
For $b=+1$ or $t>0$, $G_{f}(x,\tau)$ describes a point signal emitted by the electron  at $\tau_{ret}=0,$ and that has propagated to $x$ along the fiber $f,$ of the future lightcone of $z(\tau_{ret})$;  for $b=-1$ or $t<0,$  $G_{f}(x,\tau)$ describes a point signal that is propagating  along the fiber $\bar{f}$ of the future lightcone of $x$ towards the point $z(\tau_{adv})$ where it will be absorbed (annihilated) by the electron. See the Figure 3.
Observe the differences from the standard interpretation of the Li\`enard-Wiechert solutions. There is no advanced , causality violating solution here. $J$ is the source of the $f$ solution and a sink for the $\bar{f}$ solution. These two solutions correspond to creation and annihilation of particles, of classical photons. $G_{f}$ has no singularity, in contradistinction to the standard Green function
\be
\l{sg} 
G(x,\tau)=\frac{1}{r}\Theta(bt)\delta(r-bt),
\ee
 which is retrieved \cite{hep-th/9708066} by 
\be
\l{gg}
G(x,\tau)=\frac{1}{4\pi}\int d^{2}\Omega_{f}G_{f}(x,\tau).
\ee
So the classical photon $A_{f}$ doesn't know any singularity. The singularity presented by its average $A(x)$ is a consequence of this averaging process on its definition (\ref{gg}). The consequences and the meaning of (\ref{sg}) and  (\ref{gg}) shall be made clearer  after equation (\ref{gg2}) and we will return to discuss them at the conclusion of this note.\\

\vglue2cm

\parbox[]{7.5cm}{
\begin{figure}
\vglue-10cm
\epsfxsize=400pt
\epsfbox{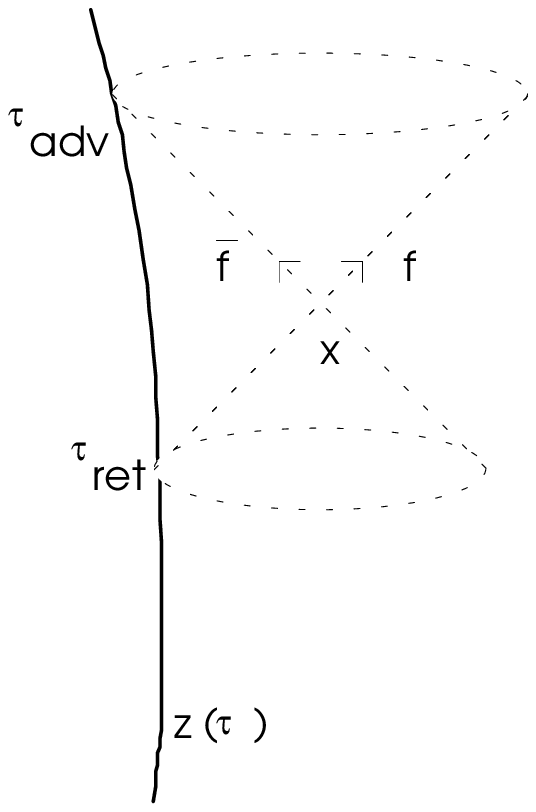}
\vglue-5cm
\end{figure}
\vglue-5cm}\\
\mbox{}
\hfill
\hspace{5.0cm}
\parbox[]{7.5cm}{\vglue-2.5cm Fig. 3. Creation and annihilation of classical particles as solutions of the wave equation. At the event $x$ there are two photons: one, on the fiber $f$, was created by the charge at $\tau_{ret}$, and the other one at the fiber $\bar{f}$  will be annihilated by the charge at $\tau_{adv}$. They are both retarded solutions.
}\\ \mbox{}
\vglue2cm

\begin{center}
The photon field
\end{center}

Let us now apply this $f-$formalism to the electromagnetic field generated by a classical point spinless electron. In this formalism where $\tau$ is treated as an independent fifth parameter, a definition of a four-vector current must carry an additional constraint expressing the causal relationship between two events $y$ and $z$. Its four-vector current is given by
\be
\l{J}
J^{\mu}(y,\tau_{y})= eV^{\mu}(\tau_{z})\delta^{3}({\vec y}-{\vec z})\delta(\tau_{y}-\tau_{z}),
\ee
where $z^{\mu}(\tau_{z}),$ is the electron worldline parameterized by its proper-time $\tau_{z}.$ The sub-indices indicate their respective events. For $b=+1$, that is, for the field emmited by J we have
\be
A_{f}(x,\tau)=2e\int d^{5}y G_{f}(x-y)V^{\mu}(\tau_{y})\delta^{3}({\vec x}-{\vec y})\delta(\tau_{x}-\tau_{y}),
\ee
where the factor 2 accounts for a change of normalization with respect to (\ref{sgf}) as we are now excluding the annihilated photon (the integration over the future lightcone). Then,
\be
\l{Af1}
A_{f}(x,\tau)=eV^{\mu}(\tau_{z})\theta[-{\bar f}.(x-z)]\theta[-f.(x-z)]{\Big |}_{{\tau_{z}=\tau_{x}+f.(x-z)}}.
\ee
So, the field $A_{f}$ is given, essentially, by the charge times its four-velocity at its retarded time.\\ 
The Maxwell field $F_{f}$ at the fiber $f$ is defined to be  $F^{f}_{\mu\nu}=(\nabla_{\mu}A^{f}_{\nu}-\nabla_{\nu}A^{f}_{\mu})$, but let us show now that $\nabla\theta(f.\Delta x)$ and $\nabla\theta({\bar f}.\Delta x)$ do not contribute to $\nabla A_{f},$ except at $\Delta x=x-z=0.$ 
The photon, being a massless field, propagates without changing its propertime, $\Delta\tau=f.\Delta x=0$; but this condition must be implemented only after the calculation is done.
 The derivation of $\theta({\bar f}.\Delta x)$ generates a $\delta({\bar f}.\Delta x)$ which with $f.\Delta x=0$ requires $\Delta x=0$, and $\nabla\theta(f.\Delta x)=\delta(f.\Delta x)f_{\nu}(1+f.V)=0$ as $f.V=-1$.\\
For notation simplicity we take $\tau_{ret}=0$ and $z(\tau_{ret}=0)$, so that we can replace $\Delta\tau$ and $\Delta x$ by $\tau$ and $x$, respectively; and so, with $x\ne0$ and $t>0$ we write just

\be
\l{dA}
\partial_{\nu}A^{\mu}_{f}=\partial_{\nu}(eV^{\mu}){\Big |}_{\tau+  f.x=0}=ef_{\nu}a^{\mu}{\Big |}_{\tau+  f.x=0},
\ee
with $a^{\mu}=\frac{dV^{\mu}}{d\tau}.$ 
So, we have
\be
\l{FedA}
F^{f}_{\mu\nu}=-e(f_{\mu}a_{\nu}-f_{\nu}a_{\mu}){\Big |}_{\tau+  f.x=0}
\ee
$F_{f}$ describes the electromagnetic field of a classical photon $A_{f}$:
\be
\l{Ef}
E_{f}^{i}=-F^{4i}_{f}{\Big |}_{\tau+  f.x=0}=e(f^{4}a^{i}-f^{i}a^{4}){\Big |}_{\tau+  f.x=0},
\ee
and 
\be
\l{Bf}
B^{i}_{f}=-e\epsilon_{ijk}F^{jk}_{f}{\Big |}_{\tau+  f.x=0}=e\epsilon_{ijk}f^{j}a^{k}{\Big |}_{\tau+  f.x=0}.
\ee
Then we see  that ${\vec E}_{f},\;{\vec a}$ and ${\vec f}$ belong to a same plane, that ${\vec B}_{f}$ is by definition orthogonal to ${\vec f}$, and therefore, that ${\vec E}_{f}$ and ${\vec B}_{f}$ are orthogonal to each other:
$${\vec f}.{\vec B}_{f}=0$$
$${\vec E}_{f}.{\vec B}_{f}=0$$
The $f-$Poynting vector $S{_f}$ is defined by (without the standard $\frac{1}{4\pi}$ factor, which is associated to its average ${\vec S}.$)
\be
\l{pv}
{\vec S}_{f}={\vec E}_{f}\times{\vec B}_{f}= e^{2}\{{\vec f}(f^{4}{\vec a}.{\vec a}-a^{4}{\vec a}.{\vec f})-{\vec a}(f^{4}{\vec a}.{\vec f}-{\vec f}.{\vec f}a^{4})\}{\Big |}_{\tau+  f.x=0}.
\ee

The Lorentz gauge condition $\nabla.A_{f}=0,$  which is necessary for reducing $\nabla_{\nu}F_{f}^{\mu\nu}$ to $\nabla_{\nu}\nabla^{\nu}A_{f}^{\mu}$, implies on
\be
\l{dA0}
a.f{\Big |}_{\tau+  f.x=0}=0,
\ee
which assures that ${\vec E}_{f}$ is orthogonal to ${\vec f},$
$${\vec f}.{\vec E}_{f}=e(f^{4}{\vec a}.{\vec f}-a^{4}{\vec f}.{\vec f}){\Big |}_{\tau+  f.x=0}=ef^{4}a.f{\Big |}_{\tau+  f.x=0}=0.$$  ${\vec E}_{f},\;{\vec B}_{f}$ and ${\vec f}$ form a triad of orthogonal vectors.\\
The $f$-Poynting vector (\ref{pv}) is reduced to
\be
\l{PV}
{\vec S}_{f}={\vec E}_{f}\times {\vec B}_{f}={\vec f}e^{2}f^{4}a^{2} {\Big |}_{\tau+  f.x=0},
\ee
which confirms the physical interpretation of $A_{f}$ as a classical photon propagating along the fiber $f$.

\begin{center}
{The photon angular distribution}
\end{center}

The Lorentz condition (\ref{dA0}) restricts the possible direction of emission of a classical photon by an accelerated charge: ${\vec f}$ is orthogonal to ${\vec a}$ in the charge stantaneous rest-frame:
\be
{\vec a}.{\vec f}{\Big |}_{{{\vec V}=0}\atop{\tau+  f.x=0}}=0,
\ee
as $a.V=0$ requires $a^{4}{\Big |}_{{{\vec V}=0}\atop{\tau+  f.x=0}}=0.$
If the electron motion is such that ${\vec V}$ and ${\vec a}$ are colinear vectors, then with a boost along ${\vec V}$, we have for the angle $\theta$ between ${\vec V}$ and ${\vec f}$:
$$\tan\theta=\frac{\sin\theta'}{\gamma(\beta+\cos\theta')}=\frac{1}{\gamma\beta},$$ or $$\sin\theta=\sqrt{1-\beta^{2}},$$ where $\beta= |\frac{d{\vec z}}{dt_{z}}|.$ So, a ultra-relativistic electron emmits a photon in a very narrow cone about the ${\vec V}$ direction. The direction of photon emmission coincides with the direction of maximum emmission in the standard formalism \cite{Jackson,Ternov}. See the Figure 4.

\hspace{-4.0cm}
\parbox[]{7.5cm}{\hspace{-2cm}
\begin{figure}
\vglue-8cm
\hspace{1.0cm}
\epsfxsize=400pt
\epsfbox{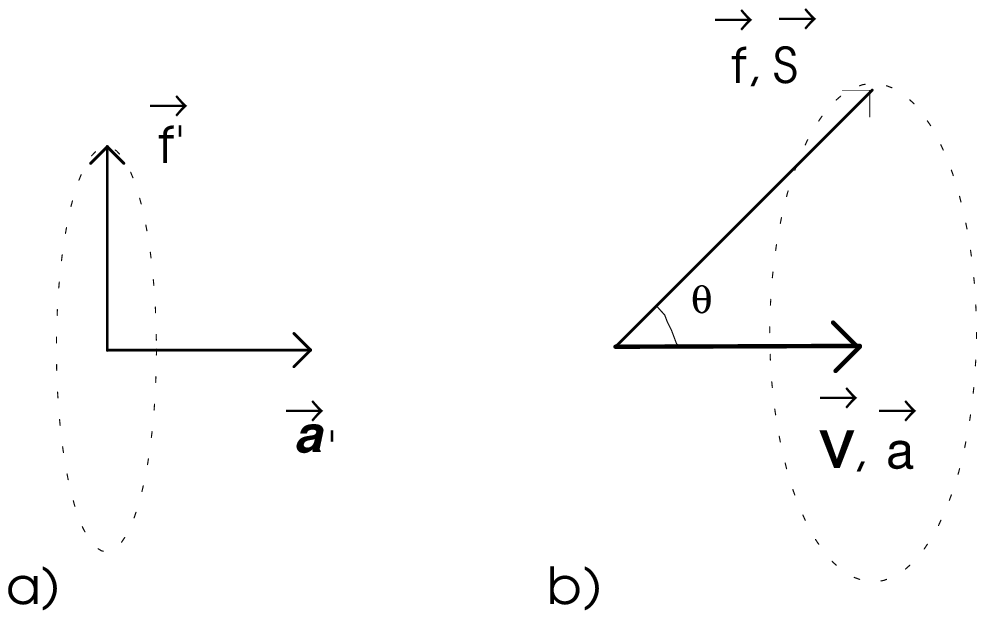}
\vglue-6cm
\end{figure}
\vglue-8cm
}
{}\\
\mbox{}
\vglue1cm
\hfill
\hspace{6.0cm}
\parbox[]{7.5cm}{Fig. 4. Radiation from an electron in a straight-line motion. The Lorentz condition requires that ${\vec f}'$ be orthogonal to the electron three-vector acceleration ${\vec a}'$ on its instantaneous rest frame: ${\vec a}'.{\vec f}'=0.$ In the laboratory frame the radiation is restricted to a cone of aperture $\theta$, with $sin\theta=\sqrt{1-\beta^{2}},$ $\quad{\vec \beta}=\frac{d{\vec x}}{dt}$.}\\
 \mbox{}
\vglue1.0cm 
If the charge is in a circular motion then ${\vec a}.{\vec V}=0$, which from  $a.V=0$ implies also that $a^{4}=0$ and then, with (\ref{dA0}), that ${\vec a}.{\vec f}=0.$\\ 
Inverting (\ref{Bf}) we find that
\be
\l{fv}
{\vec f}=\frac{{\vec a}\times{\vec B}_{f}+e{\vec a}{\vec f}.{\vec a}}{e{\vec a}.{\vec a}}=\frac{{\vec a}\times{\vec B}_{f}}{e{\vec a}.{\vec a}}.
\ee

Then we conclude from (\ref{fv}) that, for a charge in a circular motion, ${\vec f}$ is orthogonal to the plane defined by ${\vec B}_{f}$  and ${\vec a}.$ But, in the absence of magnetic monopoles, ${\vec B}_{f}$ is always orthogonal to ${\vec V}$ and so we have that the field sincroton radiation is always emitted in the ${\vec V}$-direction. Again, this agrees with the direction of maximum emission in the standard formalism \cite{Jackson,Ternov}. The $f-$formalism is just uncomparably simpler.

\begin{center}
{The photon energy-momentum}
\end{center}

The invariant $F^{2}=F^{\mu\nu}F_{\mu\nu}$  takes the following form
\be
\l{F2}
F^{2}=-2e^{2}(a.f)^{2}{\Big |}_{\tau+  f.x=0},
\ee
and is  null as a consequence of (\ref{dA0}). Therefore, $|{\vec E}_{f}|=|{\vec B}_{f}|$ is also a consequence of the Lorentz gauge.
The electron self-field energy-momentum tensor at the fiber $f,$ defined by (also without the standard $\frac{1}{4\pi}$ factor)
$$\Theta_{f}^{\mu\nu}=F^{\mu\alpha}F_{\alpha\nu}-\frac{\eta^{\mu\nu}}{4}F^{2}$$
becomes
\be
\l{t}
\Theta_{f}^{\mu\nu}=e^{2}{\Big\{}a_{f}(a^{\mu}f^{\nu}+a^{\nu}f^{\mu})-f^{\mu}f^{\nu}a^{2}+ \frac{\eta^{\mu\nu}}{2}a_{f}^{2}{\Big\}}{\Big |}_{\tau+  f.x=0},
\ee
where $a_{f}=a.f.$
With the Lorentz condition it is reduced to
\be
\l{tr}
\Theta_{f}^{\mu\nu}=-e^{2}f^{\mu}f^{\nu}a^{2}{\Big |}_{\tau+  f.x=0}.
\ee
Its physical interpretation is quite clear: $\Theta_{f}$ is not null only at those events that satisfy the constraint $\tau+  f.x=0$ on the fiber $f$ that passes by the charge at the emission (retarded) time. It clearly represents the energy-momentum of a point ``photon"propagating along the lightcone generator f.\\
In the standard formalism the electromagnetic momentum-energy four-vector P is the flux of $\Theta$ through a hypersurface $\sigma$,
\be
\l{Ps}
P^{\mu}=-\int_{\sigma} d\sigma^{3}\Theta^{\mu\nu}n_{\nu},
\ee
where $n_{\nu}$ is a normal of the hypersurface $\sigma.$ In the case of $P_{f}$, its implicit constraints $\delta(\tau+f.x),$ $\delta(\tau)$ and $\delta^{2}({\vec x}_{\hbox{\tiny T}})$ reduce it to
\be
\l{P}
P_{f}^{\mu}=-\int_{\sigma} d^{3}\sigma\Theta_{f}^{\mu\nu}n_{\nu}= -\Theta^{\mu\nu}_{f}n_{\nu}{\Big |}_{\tau+f.x=0}=e^{2}a^{2}f^{\mu}{\Big |}_{\tau+f.x=0},
\ee
with a space-like n, defined by $n=f-V$, which, in the charge rest-frame, reduces to $n{\bigg|}_{V=0}=(0,{\hat n})$, where ${\hat n}$ is the normal of a spherical surface centred at the coordinate origin (the charge position)\cite{Teitelboim}; $f.n=1.$
and $  n.V=0$.
While (\ref{Ps}) is meaningful only for $r>0$ in order to avoid the singularity of (\ref{sg}) at $r=0$, $P_{f}$ is valid without restrictions.
It is important to underline that $P_{f}$ is not null only at a single point, the instantaneous photon position at its fiber $f.$ This confirms our interpretation of $A_{f}$ as a photon on a fiber $f.$
In a theory of discrete classical fields, $P_{f}^{\mu}=e^{2}a^{2}f^{\mu}$ is the energy-momentum  four-vector radiated by the charge $e$ under the acceleration $a$ at the instant of time $\tau=\tau_{ret}$ through the emission of a single photon $A_{f}(x,\tau)$ along the fiber $f$. See the Figure 5. It can be written as 
\be
\l{pff}
P^{\mu}_{f}=e^{2}\int d\tau a^{2}f^{\mu}\delta(\tau-\tau_{ret}),
\ee
or as 
\be
\l{pff1}
\frac{\Delta P^{\mu}_{f}}{\Delta \tau}=e^{2}a^{2}f^{\mu}{\Big|}_{\tau=\tau_{ret}},
\ee
\vspace{-5.0cm}

\hspace{-3cm}
\parbox[]{7.5cm}{\hspace{-2cm}
\begin{figure}
\vglue-3cm
\epsfxsize=400pt
\epsfbox{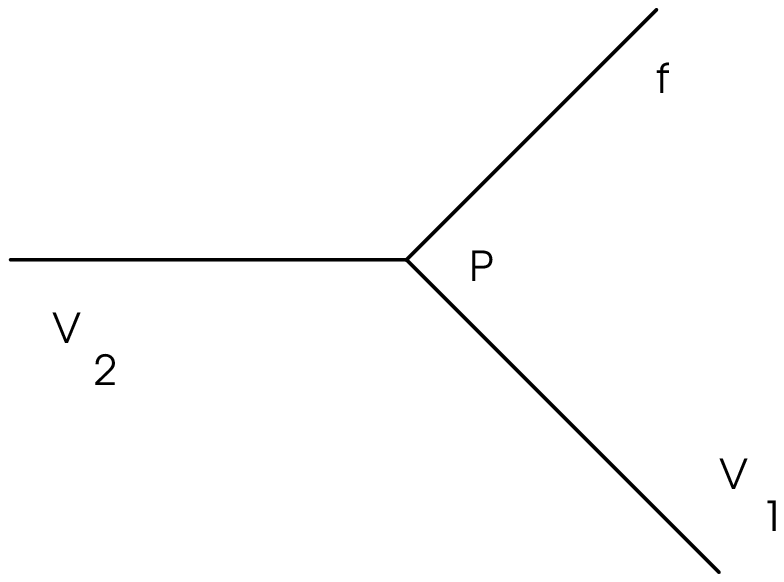}
\vglue-6cm
\end{figure}
\vglue-8cm
}
{}\\
\mbox{}
\vglue3cm
\hfill
\hspace{6.0cm}
\parbox[]{7.5cm}{Fig. 5. In the standard formalism the radiation emmission is a continuous process but in the $f-$formalism it is discrete; it just occurs at isolated events, like P on the figure, on the charge worldline $V_{1}V_{2}$.}\\ \mbox{}
\vglue.5cm 

In the standard description of a continuous emission of a continuous and distributed field $A(x,\tau)$, the discrete and instantaneously emitted $P_{f}^{\mu}$ is better replaced by $\frac{\Delta P^{\mu}}{\Delta\tau}$, and the physically real single photon $A_{f}$ must be replaced by its average (observe the return of the $\frac{1}{4\pi}$ factor).
\be
\l{Pf'}
\frac{\Delta P^{\mu}}{\Delta\tau}=\frac{1}{4\pi}\int d^{2}\Omega_{f'}P_{f'}^{\mu},
\ee
where $d^{2}\Omega_{f'}=d\phi_{f'}sin\theta_{f'}\;d\theta_{f'}$ and $\phi_{f'}$ and $\theta_{f'}$ define a generic fiber $f'(\theta_{f'},\phi_{f'})$.
This is like if $A_{f}$ has been replaced by a continuous and isotropic distribution of ficticious photons $A_{f'}$ on a hypersurface $r=const$.
What distinguishes the single, discrete and real photon $A_{f}$ from the ficticious $A_{f'}$ is that given the charge four-vector acceleration $a$ at $z(\tau_{ret})$, only $f$ satisfies the Lorentz condition $a.f=0{\big|}_{\tau=\tau_{ret}}$, as for the others $a.f'{\big|}_{\tau=\tau_{ret}}\ne0.$ So, $P_{f'}^{\mu},$ in(\ref{Pf'}), according to (\ref{t}), is not given by (\ref{P}) but by,
\be
\l{P'}
P_{f'}^{\mu}=\Theta^{\mu\nu}{\big|}_{\tau=\tau_{ret}}n_{\nu}=e^{2}\{(a_f^{2}-a^{2})f'^{\mu}+a^{\mu}a_{f}+\frac{n^{\mu}}{2}a.f'^{2}\}{\Big|}_{\tau=\tau_{ret}}.
\ee
Now, using the following well known identities \cite{Teitelboim1,Synge}
$$\frac{1}{4\pi}\int d^{2}\Omega f'^{\alpha}=V^{\alpha},$$
$$\frac{1}{4\pi}\int d^{2}\Omega f'^{\alpha}f'^{\beta}=\frac{\Delta^{\alpha\beta}}{3}+V^{\alpha}V^{\beta},$$
$$\frac{1}{4\pi}\int d^{2}\Omega f'^{\alpha}f'^{\beta}f^{\gamma}=\Delta^{(\alpha\beta}V^{\gamma)}+V^{\alpha}V^{\beta}V^{\gamma},$$
where $\Delta=\eta+VV$, and the parenthesis on the superscripts mean total symmetrization, we find from (\ref{Pf'}) and (\ref{P'}) that
\be
\l{BP}
\frac{\Delta P^{\mu}}{\Delta\tau}=\frac{2}{3}e^{2}a^{2}V^{\mu},
\ee
the standard expression of the Larmor Theorem.
The momentum-energy $e^{2}a^{2}f^{\mu}{\Big |}_{\tau+  f.x=0}$ of a single photon is then replaced in the continuum picture by its spacetime averaged value (\ref{BP}).

\begin{center}
{Physical and non-physical photons}
\end{center}

The Lorentz condition (\ref{dA0}) in this formalism has a very clear and enlightening meaning. The condition $a.f=0$ says that the photon is emitted along a lightcone generator orthogonal to the charge instantaneous four-vector acceleration. We see from (\ref{P'}) and (\ref{BP}) that in order to reproduce the radiation rate obtained from the standard continuous field $A(x,\tau)$ we necessarily must include ficticious photons $A_{f'}$ that violate the Lorentz condition. This is an indication that the average field $A(x,\tau)$ contains spurious non-physical photons. This is confirmed from an analysis of the retrieving of (\ref{gg}) from (\ref{sg}). Let us first make clearer the distinction between a physical and a non-physical field here. The physical point-field $A_{f}$ propagates along the lightcone generator $f$; $f$ is indeed its four-vector velocity. For a non-physical field $A_{f'}$ $f'$ is not co-linear to its four-vector velocity, it is not the tangent vector to its lightcone generator. This is closely related to the field satisfying or not the Lorentz condition. Now we can understand the meaning of (\ref{sg}) and (\ref{gg}). The integration over $f$, in order to retrieve the results of the standard formalism in terms of averaged fields, requires the releasing of the constraint (\ref{dA0}). So, the field $A$ is composed by the actual field $A_{f}$ that satisfies (\ref{dA0}) plus a continuous distribution of ficticious or non-physical photons that violate (\ref{dA0}) and that, consequently, has nothing to do with the physical process. This is clearly exposed in the process of retrieving the standard Green's function from $G_{f}.$ 
\be 
\l{gg1}
G(x,\tau)=\frac{1}{4\pi}\int d^{2}\Omega_{f}G(x,\tau)_{f}.
\ee
With $\tau=0$ in (\ref{pr9}) we can write $f.x=f_{4}(r\cos\theta_{f}-\varepsilon t),$ where $\varepsilon=\frac{|{\vec f}|}{f^{4}}$ and the angle $\theta_{f}$ is defined by ${\vec f}.{\vec x}=r|{\vec f}|\cos \theta_{f},\;$ for a fixed ${\vec x}$, which we take as our coordinate z-axis. This sets the inclusion of the spurious fields: the physical photons are only the ones with $\theta_{f}=0$ but the integration on $d^{2}\Omega$ is over the whole solid angle $4\pi$. Then we have
\be
\l{gg2}
G(x,\tau)=\frac{2}{4\pi}\int d\phi_{f}\sin\theta_{f} d\theta_{f}\Theta(at)\delta(r\cos\theta_{f}-\varepsilon t)=\frac{1}{r}\Theta(at)\Theta(r-\varepsilon t)=\frac{1}{r}\Theta(at)\delta(r-\varepsilon t).
\ee

The factor 2 in front the integral sign is to account for the double cone ($f^{4}>0$ and $f^{4}<0,$ the past and the future cones); we normalized $|{\vec f}|$ to 1. In the last step we made use of the constraint (\ref{1}), which implies that r cannot be larger than $\epsilon t$ and so $\theta(r-\epsilon t)$ is actually a $\delta(r-\epsilon t).$

As a final remark we observe that when the effective averaged field $A(x,\tau)$ is quantized in a manifestly covariant way we have problems with non-physical photons that, in contradistinction to the real ones, must be eliminated by the imposition of the Lorentz gauge $\partial.A=0$ as a constraint on the allowed  physical states \cite{Gupta}. This is so closely connected to our alternative approach in terms of $A_{f}$ fields (without the non-physical $A_{f'}$) that we could not avoid anticipating this comment on the quantization of the electromagnetic field, a subject to be properly discussed in a subsequent work; it justifies our conviction that one should quantize $A_{f}$ and not its average $A.$

\end{document}